\documentclass[11pt,a4paper]{article}

\pdfoutput=1
\usepackage[usenames,dvipsnames]{xcolor}
\usepackage{jheppubx}
\usepackage{enumitem}
\usepackage{graphicx}
\usepackage{dcolumn}
\usepackage{bm,psfrag}
\usepackage{overpic}
\usepackage[color=cyan!30!white,linecolor=red,textsize=footnotesize]{todonotes}

\usepackage[utf8]{inputenc}

\usepackage{amsmath}
\usepackage{cleveref}
\usepackage{subfigure}
\usepackage{float}
\usepackage[sf,isu]{caption}
\usepackage[mathscr]{euscript}

\usepackage{setspace}

\hypersetup{pdftitle={Probing thermality beyond the diagonal},    
	pdfauthor={EB,DD,SD},
	colorlinks=true,
	linkcolor=blue!70!black,
	citecolor=blue,
	filecolor=OliveGreen,
	urlcolor=RoyalBlue!60!black}

\usepackage{overpic}

\def\bra#1{{\langle}#1|}
\def\ket#1{|#1\rangle}
\def\Tr{{{\rm Tr}}}

\newcommand{\be}{\begin{equation}}
\newcommand{\ee}{\end{equation}}

\usepackage{chngcntr}

\makeatletter
\g@addto@macro\bfseries{\boldmath}
\makeatother

\newcommand{\bea}{\begin{eqnarray}}
\newcommand{\eea}{\end{eqnarray}}
\newcommand{\ba}{\begin{eqnarray}}
\newcommand{\ea}{\end{eqnarray}}
\newcommand{\nn}{\nonumber \\}

\newcommand{\beq}{\begin{equation}}
\newcommand{\eeq}{\end{equation}}
\newcommand{\beqa}{\begin{eqnarray}}
\newcommand{\eeqa}{\end{eqnarray}}
\newcommand{\beqar}{\begin{eqnarray*}}
	\newcommand{\eeqar}{\end{eqnarray*}}

\newcommand{\eg}{{\it e.g.,}\ }
\newcommand{\ie}{{\it i.e.,}\ }

\def\ausricht{\begin{aligned}}
	\def\endeausricht{\end{aligned}}

\def\tr{\rm tr}

\def\t6 {T_\mt{D6}}

\newcommand{\mt}[1]{\textrm{\tiny #1}}

\def\cale         {{\cal E}}

\def\calv         {{\cal V}}

\def\ee           {{\rm e}}

\def\tr           {\mathop{\rm Tr}}

\def\sqr#1#2{{\vcenter{\vbox{\hrule height.#2pt
				\hbox{\vrule width.#2pt height#1pt \kern#1pt
					\vrule width.#2pt}\hrule height.#2pt}}}}

\usepackage{lipsum} 

\def\b{\beta}

\def\ee{\cale}

\def\aa1{\phi}
\def\cc1{\psi}

\def\G{\Gamma}

\def\om{\Omega}

\def\ra{\Longrightarrow}
\def\vev#1{\langle #1 \rangle}

\def\eps{\epsilon}

\def\la{\langle}
\def\ra{\rangle}

\def\vev#1{\langle{#1}\rangle}

\def\pd{\partial}

\def\G{\mathcal{G}}

\def\makeitsmall{\begin{footnotesize}}
	\def\endmakeitsmall{\end{footnotesize}}

\def\nn{\nonumber}

\def\cV{\mathcal{V}}

\def\om{\omega}

\def\O{\mathbb{O}}
\def\J{\mathcal{J}}
\def\D{{\Delta}}
\def\alphaa{\alpha_{\text{avg}}}

\def\Davg{{\Delta_{\text{avg}}}}
\def\Eavg{{E_{\text{avg}}}}

\begin{document}

\title{Probing thermality beyond the diagonal}

\author[1]{Enrico M.~Brehm}
\author[1]{\!,~Diptarka Das}
\author[2]{and Shouvik Datta}

 \vspace{.1cm}

\affiliation[1]{	\vspace{0.2cm}
	Max Planck Institut f\"ur Gravitationsphysik,\\
	Albert-Einstein-Institut, \\
	Potsdam-Golm, D-14476, 
	Germany.
	\vspace{0.2cm}}

\affiliation[2]{Institut f\"{u}r Theoretische Physik,\\ 
	ETH Z\"{u}rich, \\
	Wolfgang-Pauli-Strasse 27,	8093 Z\"{u}rich, Switzerland.\\  \vspace{-0.2cm}
}

\emailAdd{$\lbrace$brehm,ddas$\rbrace$@aei.mpg.de}
\emailAdd{shdatta@ethz.ch}

\abstract{We investigate the off-diagonal sector of eigenstate thermalization using both local and non-local probes in 2-dimensional conformal field theories. A novel analysis of the asymptotics of OPE coefficients via the modular bootstrap is performed to extract the behaviour of the off-diagonal matrix elements. We also probe this sector using semi-classical heavy-light Virasoro blocks.	The results demonstrate signatures of thermality and confirms the entropic  suppression of the off-diagonal elements as necessitated  by the eigenstate thermalization hypothesis. 
 }

{
\hypersetup{urlcolor= blue!80!black}
\setlength{\parskip}{2pt}
\maketitle
\hypersetup{urlcolor=RoyalBlue!60!black}
}
\section{Introduction}

The interplay of ordering and fluctuations is important to understand the origin of statistical laws governing phases of matter. In the thermodynamic limit these laws begin to operate and control collective behaviour with an accuracy increasing with the number of degrees of freedom. Most scenarios involve a \textit{coarse-graining} that gives rise to the exact statistical laws. 
The fluctuations are inherently quantum mechanical, and as we coarse grain over increasing number of quantum states, statistical thermodynamics takes over and just a few parameters like temperature, pressure, and volume describes the system. 
This is especially true if the system possesses a finite energy density that we can measure. When this happens we say that the system under consideration has thermalized.

Thermal states in quantum mechanics  are mixed states and there is no way that unitary operations can get us from a pure state to a thermal one. Naturally, this gives rise to an observable dependent notion of thermalization, \ie one can look at the coarse grained expectation value of an operator, $\sum_\psi \vev{\psi|\O|\psi}$ and try to approximate it with a thermal expectation value $Z(\beta)^{-1}\Tr[\O\,e^{-\beta H}]$. When the operator $\O$ is arbitrary and has support over a certain subregion $V$, then the coarse-grained reduced density matrix is well approximated by the reduced thermal density matrix. Usually the quantum states under consideration will have a complicated time evolution and consequently it is interesting to understand how it approaches thermalization at late times. 

The eigenstate thermalization hypothesis (ETH) provides a mechanism (or, strictly speaking, a criterion) for closed quantum systems to be well described at late times by thermal equilibrium under unitary evolution \cite{srednicki1996thermal,Deutsch:1991,rigol2008thermalization}. The idea centers around the notion that Hamiltionian time evolution of eigenstates is trivial and, therefore, finite energy-density eigenstates are approximately thermal. 
The statement of the ETH proposes the following ansatz for the matrix elements of appropriate few-body observables
\begin{align}\label{eth-ansatze}
\bra{m} \O \ket{n} = g_{\O}(E_m)\delta_{mn} + e^{-S(\Eavg)/2} f_{\O}(\Eavg,\omega) R_{mn}.
\end{align}
Here, $\Eavg=(E_m+E_n)/2$, $S(\Eavg)$ is the thermal entropy at the average energy $\Eavg$ and $\omega=E_{m}-E_n$. The functions $g_{\O}$ and $f_{\O}$ are smooth functions of their arguments. $R_{mn}$ denotes random variables distributed with a zero mean and unit variance. The connection of this ansatz with statistical mechanics is through $g_\O(E)$ which is the statistical mechanics prediction for the operator $\O$ at the mean energy $E$. Eigenstate thermalization in this manner has been observed on the lattice for many non-integrable models, \textit{e.g.} \cite{rigol2008thermalization,Rigol2010,Biroli2010,Neuenhahn2012,Steinigeweg2013,Kim2014,Steinigeweg2014,Beugeling2015}. 
A crucial component of the above ansatz is the requirement that the off-diagonal elements are exponentially suppressed compared the diagonal ones.  The intuition behind this proposed suppression can be illustrated as follows~\cite{Rigoltalk}. Consider a system in a pure state which is a generic superposition of the energy eigenstates, $\ket{\psi} = \sum_m C_m \ket{m}$. The unitary evolution of an observable in this state is simply given by
\begin{align*}
\vev{\O(t)} 
= \sum_m |C_m|^2 \vev{m|\O|m} + \sum_{n, \, m\neq n}  C_m^* C_n e^{i (E_m -E_n)t} \vev{m|\O|n} .
\end{align*}
The late time average of the above quantity is given by the diagonal sector alone \ie $\overline{\vev{\O(t)}}= \sum_m |C_m|^2 \vev{m|\O|m}$. The fluctuations are then encoded by the off-diagonal elements
\begin{align*}
\vev{\O(t)} - \overline{\vev{\O(t)}} = \sum_{m\neq n}  C_m^* C_n e^{i (E_m -E_n)t} \vev{m|\O|n} . 
\end{align*}

\noindent
Let us make a rough estimate on these fluctuations. For an appropriately normalized pure state built from a superposition of $N$ eigenstates, we have $C_m \sim 1/\sqrt{N}$. The fluctuation is therefore
\begin{align*}
\vev{\O(t)} - \overline{\vev{\O(t)}} \sim  \sum_{m\neq n}   \frac{e^{i (E_m -E_n)t}}{ N} \vev{m|\O|n} \sim \frac{\sqrt{N^2}}{N} \vev{m|\O|n}_{\substack{\text{typical} \\ n \neq m}} \sim \vev{m|\O|n}_{\substack{\text{typical} \\ n \neq m}} . 
\end{align*}
In the second step we have used the fact that the average amplitude of $M$ random phases grows as $\sqrt{M}$ \cite{beckmann1962statistical}. This shows that the typical values of the off-diagonal elements determine the fluctuations at late times. A similar estimate shows that the typical values of diagonal elements determine the time-average $\overline{\vev{\O(t)}} \sim \vev{m|\O|m}_{\text{typical}}$. Therefore, in order for the fluctuations to be small to allow equilibriation, a necessary requirement is that the values of typical off-diagonal matrix elements should be much smaller than the diagonal ones.\footnote{ {It is worthwhile to note that there are other candidate mechanisms trying to explain thermalization unitarily. For example, there is a theorem proved in \cite{jvn} that explains thermalization of  observables in finite (but large) dimensional Hilbert spaces. A key assumption in the theorem was shown to be equivalent to the coarse-grained analogue of \eqref{eth-ansatze} in \cite{sr}; {our result fits in well within this context}. }}

This suppression is also suggested from close relations between quantum chaos and thermalization. The chaotic behaviour is effectively described by random matrix theory which also predicts the  precise factor of $e^{-S/2}$ for off-diagonal elements of hermitian operators. This suppression has been verified using numerical simulations in a number of lattice models \cite{rigol2017off} (including the SYK \cite{Sonner:2017hxc,Hunter-Jones:2017raw}). However, there has been a lack of analytical handle so far to confirm this prediction. This can be partly attributed to the lack of integrability in systems with a random matrix like behaviour. In this work, we take some steps in this direction by focussing on two dimensional conformal field theories. {In 2$d$ CFTs,   {the validity and} related consequences which follow from \eqref{eth-ansatze} have been investigated in some works \cite{Lashkari:2016vgj,Basu:2017kzo,Lashkari:2017hwq}.}  

The constraints from modular invariance allow us to extract features of both the diagonal as well as the  off-diagonal matrix elements. In the CFT parlance, the  matrix elements of local observables are nothing but the OPE coefficients.  The modular bootstrap that we shall perform uses the two-point function on the torus as the starting point.  We shall show that the off-diagonal elements, {when \textit{coarse-grained}}, are suppressed {at least} by the factor of $e^{-S/2}$, thereby  {providing evidence to} the expectation \eqref{eth-ansatze} {for a typical matrix element}. This analysis is along the lines of the recent progress on extraction of asymptotics of OPE coefficients using modular features of correlation functions \cite{km,Das:2017vej,Das:2017cnv}. We shall also utilize statistics of the OPE coefficients and the inequality between the 1-norm and the 2-norm, to provide a new constraint on the light data of the CFT. 

We also examine the off-diagonal sector using a bi-local probe of two light operators $L(x)L(y)$. This observable has been intensely explored by a variety of approaches in the context of eigenstate thermalization, holography and black holes \cite{Fitzpatrick:2014vua,Asplund:2014coa,Fitzpatrick:2015foa,Fitzpatrick:2015dlt,Fitzpatrick:2015zha,Alkalaev:2015lca,Alkalaev:2015fbw,Hijano:2015qja,Hijano:2015rla,Fitzpatrick:2016mjq,Fitzpatrick:2016mtp,Chen:2016cms,Banerjee:2016qca,Faulkner:2017hll,Kusuki:2018wcv}. We shall study the off-diagonal elements of this probe at large central charge using the monodromy method for conformal blocks. For the natural choice of perturbation parameters, in the light operator dimensions, the exchanged operator dimension and in the difference in dimensions of the heavy primaries ($H_1$ and $H_2$), the off-diagonal conformal blocks display thermal features. The inverse temperature now is given by $\beta =  L(12 \Davg/c -1)^{-1/2}$, where $\Davg=(\D_{H_1}+\D_{H_2})/2$. This is the expected temperature consistent with the ETH ansatz, similar to the diagonal case \cite{Fitzpatrick:2015foa}. This observable has also been analysed beyond the diagonal sector previously \cite{Fitzpatrick:2015zha}. As we shall show, the blocks from the monodromy method cover a different regime in the parameter space. The agreement in the large $c$ limit and deviations away from it will also be seen by comparing with blocks obtained via the Zamolodchikov recursion relations \cite{Zamolodchikov:1985wn}. A dual holographic version of these conformal blocks will also be discussed. 

The outline of this paper is as follows. In section \ref{sec:off-diag-1-point} we extract the asymptotics of  mean-squared OPE coefficients using modular bootstrap of torus 2-point functions. The asymptotics will be derived for generic states first and then refined for primaries using properties of torus blocks. In the following section \ref{sec:statistics} we discuss the statistics of our results and, in particular, use it together with previous results from \cite{km} to obtain a constraint on OPE coefficients involving light excitations. Section \ref{sec:off-diag-2-point-functions} contains the analysis of off-diagonal elements of observables $L(x)L(y)$ using the monodromy method and comparison to other approaches. We conclude in section \ref{sec:conclusions}.

\section{Off-diagonal one point functions}
\label{sec:off-diag-1-point}

\subsection{Modular properties of correlation functions on the torus}

Conformal field theories in 2-dimensions can be uniquely specified by their central charge $c$, their spectrum, \ie a set of conformal weights $\{(h_i,\bar{h}_i)\}$ of primaries $\O_i$, and coefficients $C_{ijk}$ appearing in 3-point correlation functions.
The coefficients are also directly related to the fusion coefficients $C^k_{ij}$ which tell how the field $\O_k$ contributes to the operator product expansion of  the primaries $\O_i$ and $\O_j$. Any correlation function can be constructed in terms of these data.  
Bootstrap is a strategy to gain information about the above data by using various kinds of consistency relations. The most prominent bootstrap method comes from constraints imposed by crossing symmetry of four-point correlation functions on the sphere, which has led to impressive results in higher-dimensional CFTs \cite{Rattazzi:2008pe,Rattazzi:2010yc,Vichi:2011ux,ElShowk:2012ht,Beem:2013qxa,Nakayama:2014yia}. Another type of constraint special to CFT in 2$d$ comes from modular covariance. Defining CFTs consistently on the torus implies specific transformation properties of correlation functions under modular transformations. In particular, the partition function must be modular invariant, which gives highly non-trivial constraints on the spectrum. For example, these allow the ADE classification of minimal models \cite{Cappelli:2009xj} and also imposes universality of the high energy density of states in any (unitary) 2$d$ CFT \cite{Cardy:1986ie}. 

\subsubsection*{Set-up}
Consider a torus specified by the modular parameter $\tau$. All modular transformations, \ie transformations that lead to equivalent tori, can be generated by the two transformations 
\begin{equation}
S: \tau \mapsto -\frac{1}{\tau}\,, \qquad T: \tau \mapsto \tau+1\,.
\end{equation}
The group of modular transformations is $SL(2,\mathbb{Z})$ and the most general transformation of the modular parameter is $\gamma \cdot \tau = \frac{ a\tau + b}{c \tau + d}$. We shall be interested in the correlation function of primary operators on the torus. The operator is located at the elliptic variable, $w$, which transforms to $w/(c\tau+d)$.
Therefore, primaries transform as 
\begin{align}\label{op-trans}
\O (w,\bar w) |_{\tau} &= \left[\frac{\pd (\gamma \cdot w)}{\pd w}\right]^{h}\left[\frac{\pd (\gamma \cdot \bar w)}{\pd \bar w}\right]^{\bar h} \O (\gamma \cdot w,\gamma \cdot \bar w)   \nn \\ & = (c\tau+d)^{-h} (c\bar \tau+d)^{-\bar h}\, \O (\gamma \cdot w,\gamma \cdot \bar w)\big|_\frac{a\tau + b}{ c\tau +d } .
\end{align}
Two-point correlation functions on the torus are defined as 
\begin{align}\label{two-point-def}
G(w_{12},\bar{w}_{12}|\tau)\equiv\la \O (w_1,\bar w_1)\O (w_2,\bar{w}_2)\ra_{\tau} \equiv \tr \left[ \O(w_1,\bar w_1)\O  (w_2,\bar{w}_2)q^{L_0-c/24} \bar q^{\bar L_0-c/24} \right],
\end{align}
with, $q=e^{2\pi i \tau}$. This quantity is doubly-periodic in $w_{12}$. Note that this is the unnormalized two-point function on the torus in which we do not divide by the partition function. Using the definition \eqref{two-point-def}, equation \eqref{op-trans} along with the fact that the partition function is modular invariant,  it can be seen that the correlator has the following modular transformation
\begin{align}\label{eq:modCov}
G(\gamma\cdot w_{12},\gamma\cdot \bar{w}_{12}|\gamma\cdot \tau) = (c\tau+d)^{2h} (c\bar \tau+d)^{ 2\bar h}  \, G( w_{12},  \bar{w}_{12}|  \tau).
\end{align}

\noindent
The torus, $\mathbb{T}^2 \equiv \mathbb{S}^1_\beta \times  \mathbb{S}^1_L$, describes a CFT on a spatial circle $L$ and at finite temperature $\beta$. The modular parameter $\tau = i{\beta}/L$ is purely imaginary. It follows immediately from \eqref{eq:modCov} that there is a direct relation between the high and low temperature behaviour of the 2-point functions via the S-modular transformation which takes $\tau \rightarrow -1/\tau$. In 2$d$ CFTs the low temperature behaviour is dictated by the light spectrum of the theory and allows very good approximations on various quantities. For example, in case of the partition function one can directly relate the asymptotic density of high energy states to the energy of the ground state (Casimir energy) which in turn is determined by the central charge \cite{Cardy:1986ie}. In \cite{km} a similar analysis was carried out using torus one point functions of primary operators. This yielded an average value of three point coefficients $C_{E\O E}$, where $E$ labels an operator $H$ with large dimensionless energy and $\O$ denotes a primary operator which is light compared to $H$. The average is over all states with that high energy.\footnote{The analysis was recently generalized for higher dimensional CFTs in \cite{Gobeil:2018fzy}.}

We shall now apply the modular bootstrap method on the torus two-point functions. This will lead to an average value for $C_{E\O E'}^2$ for two high energy operators $H$ at energy $E$ and $H'$ at energy $E'$, and some scalar primary operator $\O$. 

\subsection{Asymptotics of OPE coefficients}
We begin with the thermal two-point function of a primary operator $\O$. Without loss of generality, we place the operators at $(0,t)$ and $(0,0)$
\begin{align}
\la  \O(0,t)\O(0,0)  \ra_{\beta}
=\tr[ \O(0,t)\O(0,0)e^{-\beta  H }]\,,\label{eq:begin}
\end{align}
where $H = \frac{2\pi}{ L} (L_0+\bar{L}_0 -\frac{c}{12})$.
At low temperatures ($L/\beta \to 0$) the leading contribution comes from the vacuum and this becomes the two-point function on the cylinder of circumference $L$ which is completely fixed by conformal invariance
\begin{align}
\la \O(0,t)\O(0,0)\ra_{\beta} &= Z(\beta) \left(\bra{0} \O(0,t) \O(0,0)\ket{0} +\dots \right) \nn \\&= e^{\frac{\pi c \beta}{6L}}\frac{(-1)^{-\D_\O}(\tfrac{\pi}{L})^{2\D_\O}}{\sin^{2\D_\O}(\tfrac{\pi t}{L})} +\mathrm{O}(e^{-2\pi \beta \D_{\chi}/L}).
\end{align}

\noindent
Since we are considering the unnormalized correlator, we have  the factor of $Z(\beta)= e^{\frac{\pi c \beta}{6L}}$. The subleading corrections are contributions from non-vacuum states which are exponentially suppressed as $e^{-\beta \D_{\chi}}$. Here $\D_\chi$ is the conformal dimension of the lightest operator in the CFT for which the four-point function $\langle \chi \O\O\chi\rangle$ is non-zero -- this may be a primary or the lightest descendant of the vacuum. 

Using the S-modular transformation of the  two point function on the torus we are led to the high temperature result ($\beta/L \to 0$)
\begin{equation}
\la \O(0,t)\O(0,0)\ra_{\beta} = \frac{(-1)^{-\D_\O}\left(\frac{\pi}{\beta}\right)^{2\Delta_\O}}{\sinh^{2\Delta_\O}\!\left(\frac{\pi t}{\beta}\right)} e^{\frac{\pi L c}{6\beta}} + \cdots . \label{2-high}
\end{equation}
The terms in the ellipsis are suppressed in powers of $e^{-2\pi L/\beta}$. Note that in general we consider complexified time in this text, where periodicity in the real part corresponds to the thermality of the system (\ie the KMS condition) and evolution in physical time $t$ corresponds to the imaginary part\footnote{This also leads to the additional factor of $(-1)^{-\D_\O}$ in \eqref{2-high}.}. Hence, we have $\O(0,t) \equiv e^{itH} \O(0,0) e^{-itH}$. Expanding the RHS of \eqref{eq:begin} with an insertion of a complete set of states then gives ($\om_{ij}=E_i -E_j$)
\begin{align}\label{spectral}
\tr[\O (t) \O(0)e^{-\beta H}]&= \sum_{i,j} \la i | \O | j  \ra \la j  | \O  | i \ra e^{it   \frac{2\pi}{L}\om_{ij}}   e^{- \frac{2\pi\beta }{L}(E_i-\frac{c}{12})}\,. 
\end{align}
Here $E_i$ is the dimensionless energy (= conformal dimension $\D_i$) of the eigenstate $\ket{i}$ which includes all states of the CFT, both primaries and their descendants. An integral representation of the above formula is 
\begin{align}\label{int-2pt}
\tr[\O (t) \O(0)e^{-\beta H}]&= 
\int_0^{\infty} dE \int_{-\infty}^\infty d\om  \  \J_{\O}(E,\om) e^{i t  \frac{2\pi\om}{L} }  e^{-\frac{2\pi\beta}{L} (E-c/12)}\,,
\end{align}
where we have introduced the weighted spectral density ($\om =E' - E$)
\begin{align}\label{JO-def}
\ausricht
\J_{\O}(E,\om) =\J_\O(E,E')   
&\equiv\sum_{i,j} | \la i | \O  | j  \ra|^2  \delta(E_i -E)  \delta(\om_{ij}-\om) \,  ,    \\    &=  \sum_{i,j} | \la i | \O  | j  \ra|^2  \delta(E_i -E) 
\delta(E_j -E')  \, . 
\endeausricht
\end{align}

\noindent 
It is clear that \eqref{int-2pt} has the structure of a Laplace transform $\mathcal{L}$ in $\Delta$ and a Fourier transform ${\cal F}$ in $\omega$ of the weighted spectral density, \ie
\begin{equation}
\la \O(0,t)\O(0,0)\ra_{\beta} = e^{\frac{\pi c \beta}{6L}} {\cal F}\left[ {\cal L}\left[\J_{\O}(E,\om)\right](\beta)\right](t)
\end{equation}
The weighted spectral density can then be obtained by inverting this expression. We get
\begin{align}
\J_{\O}(E,\om) &=  \mathcal{L}^{-1}\mathcal{F}^{-1}\left[ e^{-\frac{\pi c \beta}{6L}}\la \O(0,t)\O(0,0)\ra_{\beta} \right](E,\om)\nn \\
& \equiv \left(\frac{2\pi}{L}\right)^{2\Delta_\O} \int_{\gamma-i\infty}^{\gamma+i\infty}\frac{d\tilde{\b}}{2\pi i}\, e^{\tilde{\beta}E} \int_{-\infty}^{\infty}\frac{d\tilde{t}}{2\pi}\,e^{-i\tilde{t}\om} \frac{(-1)^{\Delta_\O} \left(\frac{\pi}{\tilde{\beta}}\right)^{2\Delta_\O}e^{\frac{\pi^2 c}{3\tilde{\beta}}-\frac{ c \tilde{\beta}}{12}}}{\sinh^{2\Delta_\O}\!\left(\frac{\pi \tilde{t}}{\tilde{\beta}}\right)}\nn \\
&= \frac{\left({4\pi^2}/ {L}\right)^{2\Delta_\O}  }{ \Gamma(2\D_\O)} \int_{\gamma-i\infty}^{\gamma+i\infty}\frac{d\tilde{\b}}{2\pi i} e^{\frac{\pi^2 c}{3\tilde{\beta}}+\tilde{\beta} (E +\frac{\omega}{2}-\frac{ c }{12})}    \tilde{\beta}^{1-2\D_\O}  {\left|\Gamma \left(\D_\O+ i\frac{\tilde{\beta} \om}{ 2 \pi} \right)  \right|^2}\,. \label{inv-int}
\end{align}

\noindent
In the second line we have switched to the dimensionless variables, $\tilde{\beta} = 2\pi\beta/L$ and $\tilde{t} = 2\pi t/L$. The result for the Fourier transform in the third line follows from Mellin-Barnes integrals (see \eg \cite{Becker:2014jla} and appendix \ref{mellin}). 
We have taken $\omega > 0$, therefore we closed the $t$-contour in the lower half plane, hence picked the $-i\epsilon$ prescription in \eqref{inv-MB} for the retarded 2-point function. Note that the combination $E+\om/2$ appearing in \eqref{inv-int} is the average $\Eavg=(E+E')/2$. 

To perform the inverse Laplace transform we may utilize the saddle point approximation. Considering just the exponential factor in the integrand \eqref{inv-int}, we find that the saddle is located at 
\begin{align}\label{beta-saddle}
\beta_* =  \frac{L}{\sqrt{\tfrac{12}{c}(E+\frac{\om}{2}-\frac{c}{12})}}.
\end{align}
This leads to an `effective temperature' which is given in terms of the average ($E+\om/2$) of the dimensionless energies of the (off-)diagonal states. This thermal feature is exactly the same as observed earlier for the case of two point function of two light operators \cite{Fitzpatrick:2015zha}. In \cite{Fitzpatrick:2015zha} the same parameter was used to perform a uniformization transformation to calculate  (off-)diagonal heavy light blocks. 
To justify the saddle point approximation,  we will evaluate \eqref{inv-int} by keeping all the factors in the integrand. In what follows we will keep $\omega$ fixed while taking $E \rightarrow \infty$. The states with large $E$ are expected to dominate at high temperatures, $\tilde{\beta} \ll 1$. Hence setting $x = \tilde{\beta}\omega$, we can do an expansion around $x\rightarrow 0$. %
\begin{equation}
\left|\Gamma\left(\Delta_\O +i \frac{x}{2\pi}\right)\right|^2 = \sum_{n=0}^\infty b_n x^{2n}\,.
\end{equation}
Using the integral representation of the modified Bessel function
\begin{equation}\label{bessel}
I_\nu(z) = \left(\frac{z}{2}\right)^\nu \frac{1}{2\pi i} \int_{\tilde\gamma-i\infty}^{\tilde\gamma+i\infty}dt \frac{1}{t^{\nu+1}} e^{t + \frac{z^2}{4t}}
\end{equation}
one can perform the integral \eqref{inv-int} within the sum to get
\begin{align}
\J_{\O}(E,\om) &= \left(\tfrac{4\pi^2 }{L}\right)^{2\Delta_\O}\tfrac{1}{\Gamma(2\D_\O)} \left(\tfrac{1}{(2\pi)^2}\tfrac{12}{c} \left(E +\tfrac{\omega}{2}-\tfrac{c}{12}\right)\right)^{\Delta_\O-1}  \\
&\ \times\sum_n b_n \left(\tfrac{1}{(2\pi\om)^2}\tfrac{12}{c} \left(E +\tfrac{\omega}{2}-\tfrac{c}{12}\right)\right)^{-n} I_{2E-2-2n}\left(4\pi\sqrt{\tfrac{c}{12}\left(E +\tfrac{\omega}{2}-\tfrac{c}{12}\right)}\right)\,.\nn
\end{align}
Since we are considering only large $E$ asymptotics, we expand the Bessel function for large arguments. At large $z$, $I_\nu(z) \approx {e^z}/{\sqrt{2\pi z}}$, which in particular is independent of $\nu$. This allows us to resum the series to recover the $\Gamma$ functions, resulting finally in 
\begin{align}
\hspace{-.2cm}\left(\tfrac{L }{2\pi}\right)^{2\Delta_\O}\!\J_{\O}(E,\om) \simeq
&    \left(\tfrac{12}{c} \right)^{\Delta_\O-\tfrac{3}{4}}\frac{\left(\Eavg-\tfrac{c}{12}\right)^{\Delta_\O-\frac{1}{2}} }{ {\Gamma(2\D_\O)}}\rho(\Eavg) {\left|\Gamma\left(\! \Delta_\O + i \tfrac{\omega/2}{\sqrt{{12 \Eavg}/{c}-1}}\right)\right|^2}\,. \label{Jo}
\end{align}
Here $\rho(E)$ is the asymptotic density of high energy states given by Cardy's formula \cite{Cardy:1986ie}
\begin{equation}\label{eq:CardyDensity}
\rho(E) \simeq \sqrt{2}\pi \left(E - \frac{c}{12}\right)^{-3/4} e^{4\pi \sqrt{\frac{c }{12}\left(E-\frac{c}{12 }\right)}} \equiv \sqrt{2}\pi \left(E - \frac{c}{12}\right)^{-3/4} e^{S(E)}\,.
\end{equation}
Here $S(E) = {4\pi \sqrt{\frac{c }{12}\left(E-\frac{c}{12 }\right)}}$  is the entropy devoid of any logarithmic corrections. The result \eqref{Jo} agrees with the saddle point result of \eqref{inv-int}. However, since we actually work with a contour integral in the complex plane, we prefer the method presented. 
We can now provide the asymptotics of  the mean squared matrix element, $\overline{|C_{E\O E'}|^2} $. It can be computed from 
\begin{align}
\J_\O(E,E') &= \sum_{i,j} | \la i | \O  | j  \ra|^2  \delta(E_i -E) 
\delta(E_j -E')  \simeq \left(\frac{2\pi}{L}\right)^{2\Delta_\O}\rho(E)\rho(E') \,\overline{|C_{E\O E'}|^2} \,,
\end{align}
where we can use \eqref{eq:CardyDensity} only for both $E$ and $E'$ large. The average here is over all heavy states of the CFT and, we reiterate  that it does not distinguish between primaries and descendants. Therefore, for the leading order approximation, the mean square OPE coefficient can be written as
\begin{align}
\overline{|C_{E\O E'}|^2} &\simeq \left(\frac{12}{c} \right)^{\Delta_\O-\tfrac{3}{4}}\frac{\left(\Eavg-\tfrac{c}{12}\right)^{\Delta_\O-\frac{1}{2}} }{ {\Gamma(2\D_\O)}}\frac{\rho(\Eavg)}{\rho(E)\rho(E')} {\left|\Gamma\left(\! \Delta_\O + i \frac{\omega/2}{\sqrt{{12 \Eavg}/{c}-1}}\right)\right|^2},\,\nn  \\
&\simeq \mathcal{N}_{\O} \,  e^{-S(\Eavg)}\, \left(\frac{12\Eavg}{c}-1\right)^{\Delta_\O+\frac{1}{4}} \  \left|\Gamma\left( \Delta_\O + i \frac{\omega/2}{\sqrt{{12 \Eavg}/{c}-1}}\right)\right|^2. \label{mean-sqr}
\end{align} 
with  $\mathcal{N}_{\O} = \left(\frac{c}{12}\right)/{\sqrt2 \pi\, \Gamma(2\Delta_\O)}$. In the final step we have worked in the approximation $|\omega|=|E-E'|$ is much smaller than $\Eavg$ and also omitted all sub-leading contributions for large $\Eavg$. This approximation is justified since the $\Gamma$-function factor is peaked around $\om=0$.  The factor $\left({12\Eavg}/{c}-1\right)^{\Delta_\O+{1}/{4}}$ above can be regarded as a logarithmic correction to the entropy in presence of the probe operators.
The result demonstrates that mean square OPE coefficients of the above kind are entropically suppressed as $e^{-S(\Eavg)}$. The above equation \eqref{mean-sqr} is one of the main results of this work.

It is worthwhile to interpret the result \eqref{mean-sqr} holographically. The quantity $\overline{|C_{E\O E'}|^2}$ provides a measure of the transition rate of a black hole microstate of energy $E'$ to another of energy $E$ along with the emission of a scalar (dual to the primary $\O$). The factor of $e^{-S}$ is the probability of choosing a single black hole microstate and the $\Gamma$-function factor is the black hole emission rate for scalars \cite{Maldacena:1997ih}\footnote{In the limit of a large central charge and under the assumption that ETH holds, the 3-point coefficient can also be derived from other holographic methods \cite{Leiden}.}.

\subsubsection*{Fluctuations and eigenstate thermalization}

A measure of fluctuations of the operator $\O$ in the eigenstate $\ket{n}$ can be obtained from the following quantity; see  \cite[\S 6]{rigolReview} for further details. 
\begin{align}
\ausricht
	C_\O (t) &\equiv \bra{n} \O(t) \O(0) \ket{n} - \bra{n} \O(t) \ket{n} \bra{n} \O(0) \ket{n} =\sum_{m\neq n} |\bra{n} \O \ket{m}|^2 e^{i\om_{nm}t},  \\ 
	& \supset \sum_{m\neq n} e^{i\om_{nm}t} e^{-S(E_m+\om/2)} |f_\O (E_m+\om/2, \om )|^2 |R_{mn}|^2 .\label{r-eth-2} 
	\endeausricht
\end{align}

\noindent
The final relation is true only for states which obey the ETH ansatz \eqref{eth-ansatze}. 
Moreover, the function $f_\O(\Eavg, \om )$ is peaked around $\om=0$ and thereby the dominant contribution to sum in \eqref{r-eth-2} is from the small window of states within the regime $E_m - E_n \ll 1$. 

The form of the summand  \eqref{r-eth-2} is clearly consistent with \eqref{mean-sqr} obtained using modular bootstrap. 
Concretely, for $\ket{n}$ being a typical high energy eigenstate we can plug in the asymptotic mean square average \eqref{mean-sqr} for $|\bra{n} \O \ket{m}|^2$ in  the first line of  \eqref{r-eth-2}.  We obtain the final equality which is also predicted by ETH, \ie the final relation of \eqref{r-eth-2}. Notably, the entropic suppression $e^{-S(E_m+\om/2)}$ comes out precisely as expected. The average of the factor $|R_{mn}|^2$ is a constant.
Furthermore, we can read off the function $f_\O(\Eavg, \om )$ 
\begin{align}
|f_\O (\Eavg, \om )|^2 = \mathcal{N}_\O\left|\Gamma \left(\D_\O+ i  \frac{  \om /2}{\sqrt{12\Eavg /c -1}} \right) \right|^2 .
\end{align}
This is an explicit verification of smoothness of the function $f_\O$ as conjectured by the ETH ansatz \eqref{eth-ansatze}. It also agrees with the prediction that this is a function in the average energy of the in and out states $\Eavg$ and the frequency $\om$ characterizing the off-diagonal elements. 
It is also crucial to observe that the distribution of the matrix elements with respect to $\om$ is {\emph {not}} that of a generic Gaussian ensemble. The distribution is actually that of   generalized hyperbolic  secants \cite{harkness1968generalized}. This should be contrasted with the cases where the underlying behaviour is believed to be governed by random matrices \cite{Nation1991,rigol2008thermalization,Cotler:2016fpe,rigol2017off}. 

\subsection{Refining asymptotics for primaries}
The formula \eqref{mean-sqr} is for the averaged matrix elements of high energy eigenstates regardless of them being primaries or descendants. In this subsection,  we show that it can be refined using the torus  2-point blocks leading to a derivation of the asymptotics of the actual OPE coefficients indexed by primaries. The analysis covers $c>1$ theories with   Virasoro symmetry. 

\begin{figure}[t!]	
  \centering
  \includegraphics[width=14cm]{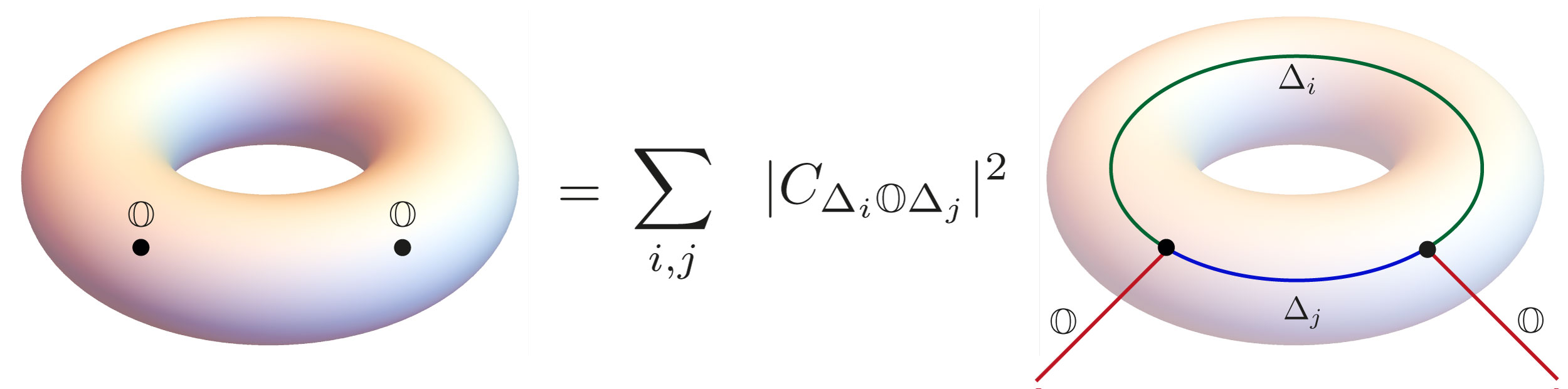}
  \caption{Schematic representation of the decomposition of the torus 2-point function in projection/necklace channel using torus blocks.}\label{necklace} 
\end{figure}

The thermal Euclidean two point function of light primaries $\O$ located at elliptic coordinates $(w,\bar{w})$ and $(0,0)$ on the torus $\mathbb{T}^2$ admits the following expansion, involving a sum only over the exchanged primaries 
\begin{align} \label{tor2}
\vev{ \O(w,\bar{w}) \O(0)}_{\mathbb{T}^2} = (2\pi)^{2\D_\O} \sum_{i,j} |C_{\D_i\O\D_j}|^2 q^{h_i-c/24}\bar{q}^{\bar{h}_i-c/24} z^{h_{ij}} \bar{z}^{\bar{h}_{ij}} \G_{ij}^\O(z,q)\bar\G_{ij}^\O(\bar{z},\bar{q}).
\end{align}
Here, $q=e^{2\pi i \tau}$, $z=e^{2\pi i w}$ and $h_{ij}=h_i-h_j$.  The factor $\G^\O_{ij}(z,q)$ and its antiholomorphic counterpart are the 2-point torus blocks which encode the contributions of the descendants of primaries. Note that we have considered the expansion along the projection channel as illustrated in Fig.~\ref{necklace}. This channel is also referred to the $s$-channel or the necklace channel literature. If we had chosen to work with the other channel (called the $t$-channel or the OPE channel) coefficients of the kind $|C_{\O\O\D_i}|^2$ would have appeared in the sum above.
As in the previous subsections, we shall work with the rectangular torus such that $q = e^{-\beta}$ (setting $L=2\pi$ for convenience) and the locations of the operators are at $(0,t_E)$ and $(0,0)$. 

The 2-point torus blocks in the projection channel have the following form \cite{Brustein:1988vb, Alkalaev:2017bzx}
\begin{align}\label{torus-block-def}
\G^\O_{ij} (q,z)= \sum_{n,m=0}^{\infty} q^n \sum_{\substack{|M|=|N|=n \\ |S|=|T|=m }} B_i^{M|N}  \frac{\vev{\D_i,M | \O(z)|S,\D_j}}{\vev{\D_i  | \O(z)| \D_j}} B_j^{S|T}  \frac{\vev{\D_j,T | \O(1)|N,\D_i}}{\vev{\D_j  | \O(1)| \D_i}},
\end{align}
and similarly for the anti-holomorphic part. $B^{P|Q}_{i,j}$ appearing above are the elements of the inverse Gram matrices of the exchanged primaries $\D_{i,j}$. All the inner products in the torus block \eqref{torus-block-def} are completely determined by the Virasoro Ward identities. 

We are interested in the high temperature regime of the two-point function \eqref{tor2} and we  specialize to the situation where the operators $\O$ are light. In this regime, the dominant contribution arises from heavy primaries. We shall also assume $h_{ij}\ll h_i$. The torus blocks for the heavy exchanged primaries have a greatly simplified form. In order to see this, note that the elements of the inverse Gram matrices in the regime $h_{i},h_j\gg 1$ always have a leading behaviour as some inverse power of $h_i,h_j$. On the other hand, the inner products in the numerator, $\vev{\D_i,P | \O(z)|Q,\D_j}$, behave as some positive power of $h_i$ for $h_i \gg h_{\O}$ \ \footnote{This is the reason why we need $\O$ to be light. This is unlike the previous subsection where there was no such restriction.} and $h_i \gg h_{ij}$. It can be seen that only if $M=S$ and $T=N$ in both inner products in \eqref{torus-block-def}, we get an order one contribution from the summand. Otherwise, these get suppressed by inverse powers of $h_{i},h_j$. The net effect due to the presence of the inverse Gram matrices is, therefore, to restrict the dominant contribution to the `diagonal sector' only
\begin{align}\label{heavy-torus-block-1}
\lim\limits_{h_{i,j}\to \infty}\G^\O_{ij} (q,z) \simeq \sum_{n}^{\infty} q^n \sum_{\substack{ |N|=n }} B_i^{N|N}  \frac{\vev{\D_i,N| \O(z)|N,\D_j}}{\vev{\D_i  | \O(z)| \D_j}} B_j^{N|T}  \frac{\vev{\D_j,N | \O(1)|N,\D_i}}{\vev{\D_j  | \O(1)| \D_i}}.
\end{align}
The sum over $N$ is over all possible descendants at each level $n$ which   can be uniquely labelled by partitions of integers. Retaining just the leading terms (which are order-one) in the $h_{i},h_j\to \infty$ limit, merely counts the number of descendants at each level  
\begin{align}\label{htb}
\lim\limits_{h_{i},h_j\to \infty}\G^\O_{ij} (q,z) \simeq \sum_n p(n)q^n = \frac{q^{1/24} }{ \eta(q)}. 
\end{align}
This is nothing but the character of the primary $h_{i}$ (together with the prefactor accounted for in \eqref{tor2}). To summarize in words,   the heavy-light regime of the torus blocks can be approximated by the character of the heavy primary, \ie the deformations caused by the presence of the light operators are negligible. This is analogous to the situation for the lower-point torus block in \cite{km,Hadasz:2009db,Alkalaev:2016ptm}. This can be checked to a very low order in the $q$-expansion using the results of \cite[Appendix A]{Alkalaev:2017bzx}. 

Using the S-modular transformation of the Dedekind-eta function, $\eta(-\frac{1}{\tau}) = \sqrt{-i\tau}\, \eta(\tau)$, we arrive at the high temperature ($\beta \rightarrow 0$) behaviour of the following product
\begin{align}
\G_{ij}^\O(z,q)\G_{ij}^\O(\bar{z},\bar{q}) \simeq \beta \exp\left( -\frac{1}{12}\beta + \frac{\pi^2}{3 \beta } \right). \label{modG}
\end{align}
The corrections to the above are in inverse powers of $h_{i,j}$ and additional exponentially suppressed corrections in $\tilde{q}=e^{-4\pi^2/\beta}$.
We can now in a position to define the weighted spectral density over primaries 
\begin{equation}
\J^P_{\O}(\Delta,\om) = \sum _{\D_i, \D_{j}} |C_{\D_i \O \D_j}|^2 \delta(\D_i - \D) \delta(\D_j-\D_i -\om)\, . 
\end{equation}
As before, the high temperature limit of the Euclidean thermal two point function is 
\begin{equation}
\la \O(0,t_E)\O(0,0)\ra_{\beta} =  \frac{\left(\frac{\pi}{\beta}\right)^{2\Delta_\O}}{\sin^{2\Delta_\O}\!\left(\frac{\pi t_E}{\beta}\right)} e^{\frac{\pi^2 c}{3\beta}}\,.   \label{euc2}
\end{equation} 
Using \eqref{modG} and \eqref{euc2} and the high-energy/high-temperature version of  \eqref{tor2} we have
\begin{align}
\frac{ \beta^{-2{\D_\O}}} {\sin^{2\Delta_\O} \frac{\pi t_E}{\beta} } e^{\frac{\pi^2 c}{3\beta} } = 2^{2\Delta_\O} \int_0^\infty d\D \int_{-\infty}^\infty d\omega ~ \J^P_{\O}(\Delta,\om)\,\beta\,   e^{-\beta ( \D - \frac{c-1}{12} )} e^{\frac{\pi^2 } {3\beta}} e^{t_E \omega}\,. 
\end{align}
Next we analytically continue both sides to Lorentzian time and follow the same steps as in the previous section to solve for $\J_\O^P$ by performing an inverse Fourier transform followed by the inverse Laplace transform. We keep only the leading order contributions for high energies. Our final result for the weighted spectral density of heavy primaries then is (with $\D_\text{avg} = \D +\frac{\omega}{2}$)
\begin{align}
\J^P_{\O}(\D_\text{avg})
&\simeq \  \frac{\left(\Delta_\text{avg}-\tfrac{c-1}{12}\right)^{\Delta_\O-\frac12}}{\sqrt{2}\pi(2\pi)^{2\D_\O}\Gamma(2\D_\O)} \left(\frac{12}{c-1} \right)^{\D_\O-\frac14} \rho^P\!\left(\D_\text{avg}\right) {\left| \Gamma \left(   \Delta_\O + i \frac{\omega/2}{\sqrt{\frac{12}{c-1}\D_\text{avg}-1}} \right) \right|^2}{}\, .  \nonumber
\end{align}
We have used the density of heavy primaries \cite{km}
\begin{align}
\rho^P(\D) \simeq 
\sqrt{2}\pi \left(\D-\tfrac{c-1}{12}\right)^{-\frac14} e^{4\pi \sqrt{\tfrac{c-1}{12}\left(\D-\tfrac{c-1}{12}\right)}} \equiv \sqrt{2}\pi \left(\D-\tfrac{c-1}{12}\right)^{-\frac14} e^{S^P(\D)}.
\end{align}
Using the latter we can also write
\begin{equation}
\J_\O^P(\D,\D') = \rho^P(\D)\rho^P(\D')\overline{|C_{\D\O\D'}|^2}\,, 
\end{equation}
where $\overline{|C_{\D\O\D'}|^2}$ now is the mean squared OPE coefficients with the average taken over all heavy primaries with conformal dimension $\D$ and $\D'$. The final result of this section therefore is 
\begin{align}
\overline{|C_{\D\O\D'}|^2} &\simeq   \frac{\left(\Delta_\text{avg}-\frac{c-1}{12}\right)^{\Delta_\O-\frac12}}{\sqrt2\pi(2\pi)^{2\D_\O}\Gamma(2\D_\O)}\! \left(\frac{12}{c\!-\!1} \right)^{\D_\O-\frac14}\! \frac{\rho^P\!\left(\D_\text{avg}\right)}{\rho^P(\D)\rho^P(\D')} {\left| \Gamma \left(   \Delta_\O + i \frac{\omega/2}{\sqrt{\frac{12}{c-1}\D_\text{avg}-1}} \right) \right|^2}{}\nn \\
&\simeq   \mathcal{N}^P_{\O} \ e^{-S^P(\Davg)} \ \left(\frac{12 \Delta_\text{avg}}{c\!-\!1}-1\right)^{\Delta_\O-\frac14} \left| \Gamma \left(   \Delta_\O + i \frac{\omega/2}{\sqrt{\frac{12}{c-1}\D_\text{avg}-1}} \right) \right|^2\label{mean-sqr-p}
\end{align}
where again the last line is true when $|\omega|$ is much smaller than $\Davg$ and we define $\mathcal{N}^P_{\O} = {\left(2\pi^2(2\pi)^{2\D_\O} \Gamma(2\D_\O)\right)^{-1}} $. The form of the above average is similar to that of \eqref{mean-sqr}. The precise shift in the central charge from $c$ to $c-1$ has also been observed in mean heavy-heavy-light OPE coefficient in \cite{km}. This is possibly related to the fact that the torus block \eqref{htb} in the heavy limit (or the characters of primaries in $c>1$ Virasoro CFTs, which count the descendants) is the same as the partition function of the $c=1$ free boson theory. It would be interesting to derive the above result holographically building on some of the techniques of \cite{Kraus:2017ezw,Alkalaev:2016ptm,Alkalaev:2017bzx}.

\section{A bound from the statistics of OPE coefficients}\label{sec:statistics}

The results of the previous section using modular bootstrap provide statistics of high energy CFT data. The expectation values of a randomly chosen probe are largely dependent on the actual distribution of the quantity of interest. A good measure for this is the squared variance $\sigma^2 = \langle X^2 \rangle - \langle X\rangle^2$, which quantify how peaked a distribution on a random variable $X$ is around its expectation value $\langle X\rangle$. The squared variance is always positive, \textit{s.t.} it follows immediately that our result on the mean squared OPE coefficient sets an upper bound for the mean OPE coefficients of \cite{km} 
\begin{align}\label{eq:ineq1}
\overline{\bra{E}\O\ket{E'}} \equiv \overline{C_{E\O E'} } \leq \overline{|C_{E\O E'}| } \leq \sqrt{\overline{|C_{E\O E'}|^2 }}\,. 
\end{align}
Note that the average here is over two sets of heavy eigenstates which are specified by $E$ and $E'$. Since $\overline{|C_{E\O E'}|^2}$ is entropically suppressed, so is the variance. The distribution of OPE coefficients $C_{E\O E'}$ is therefore rather sharply peaked around its mean value. This justifies the notion of a typical ($\sim$ almost all) OPE coefficient for which one gets
\begin{equation}
|C^\text{typ}_{E\O E'}| \simeq \overline{|C_{E\O E'}|} \leq \sqrt{\overline{|C_{E\O E'}|^2 }} \simeq e^{-\frac{S(\Eavg)}{2}} \,.
\end{equation}

\noindent
This shows that in a 2$d$ CFT typical off-diagonal OPE coefficients of high energy eigenstates are \textit{at least} suppressed as expected from ETH ansatz \eqref{eth-ansatze}. Our considerations here, however, do not allow provide a lower bound. 

If we consider two different heavy states but with the same energy, \textit{i.e.} $\omega = 0$, the mean squared OPE coefficient takes the form
\begin{equation}
\overline{|C_{\O_E\O\O'_E}|^2} \simeq \frac{1}{\rho(E)}\left(\frac{12}{c}\right)^{\D_\O-\frac{3}{4}} \left(E-\frac{c}{12}\right)^{\Delta_\O-\frac{1}{2}} \frac{\Gamma\left( \Delta_\O \right)^2}{\Gamma(2\Delta_\O)}\,.
\end{equation}

\noindent 
Note that the average here includes off-diagonal OPE coefficients with different states but having the same energy. This situation is possible since the heavy spectrum is highly degenerate. To make this clear we shall slightly change the notation. The average coming from purely diagonal entries (\ie when both the energy and the state are the same)
\begin{equation}
\sum_i |\bra{i}\O\ket{i}|^2 \delta(E_i-E) \simeq \rho(E)\overline{|C_{E\O E}|^2}\,,
\end{equation}
is bounded from above as\footnote{This is because $\rho(E) \overline{|C_{\O_E\O\O'_E}|^2}$ contains more states/terms than $\overline{|C_{E\O E}|^2}$ -- namely, the off-diagonal ones corresponding to different states having the same energy.}
\begin{equation}
\overline{|C_{E\O E}|^2}\  \le  \ \rho(E) \overline{|C_{\O_E\O\O'_E}|^2} \ \simeq \  \left(\frac{12}{c}\right)^{\D_\O-\frac{3}{4}} \left(E-\frac{c}{12}\right)^{\Delta_\O-\frac{1}{2}} \frac{\Gamma\left( \Delta_\O \right)^2}{\Gamma(2\Delta_\O)}\,. \label{eq:ineq2}
\end{equation}
This upper bound is in particular not entropically suppressed. In the diagonal case (energy and state) the \textit{mean} OPE coefficient was computed in \cite{km} from the modular properties of one point functions on the torus. The result is
\begin{equation}\label{eq:Kraus}
\overline{C_{E\O E}} \simeq \frac{C_{\chi\O\chi}}{\left(\frac{c}{12}-\Delta_\chi\right)^{\frac{\D_\O}{2}-\frac14}}  \left(E-\frac{c}{12}\right)^\frac{\Delta_\O}{2} e^{-4\pi \left(1-\sqrt{1-\frac{12\Delta_\chi}{c}}\right)\sqrt{\frac{c}{12}\left(E -\frac{c}{12}\right)}}\,,
\end{equation}
where $\chi$ is the lightest field in the spectrum of the CFT with $C_{\chi\O\chi}\neq0$. Note that also in this case \eqref{eq:ineq1} must hold which together with \eqref{eq:ineq2} is consistent with all previous assumptions made in this text and in \cite{km}, in particular with $E\gg 1$ and the unitarity bound, $\D_\chi>0$. We can even go further and constrain $C_{\chi\O\chi}$ when $\D_\chi$ is very small. Using \eqref{eq:ineq1} with \eqref{eq:ineq2} we arrive at the following inequality
\begin{equation}\label{37}
C_{\chi\O\chi}^2 \le \left(\frac{12}{c}\right)^{-\frac{1}{4}}  \left(1-\frac{12\D_\chi}{c}\right)^{\D_O-\frac{1}{2}}\frac{\Gamma\left( \Delta_\O \right)^2}{\Gamma(2\Delta_\O)} \left(E-\frac{c}{12}\right)^{-\frac{1}{2}} e^{8\pi \left(1-\sqrt{1-\frac{12\Delta_\chi}{c}}\right)\sqrt{\frac{c}{12}\left(E -\frac{c}{12}\right)}}\,.
\end{equation}

\noindent 
On the RHS the central charge $c$ and the operator dimension $\D_{\mathbb
{O}}$ are fixed by choice, and $\D_\chi$ is fixed by the assumption that $\chi$ is the lightest field in the spectrum of the CFT that  has non-vanishing OPE coefficient $C_{\chi\O\chi}$. The inequality is true for any large enough $E$. In fact, the RHS has a minimum at  
\begin{equation}
E_\text{min} = \frac{c}{12} \frac{1-\frac{6\D_\chi}{c}+\sqrt{1-\frac{12 \D_\chi}{c}}+32\pi^2\D_\chi^2}{32\pi^2\D_\chi^2}\,
\end{equation}
which is large for $\D_\chi^2\ll \frac{c}{192\pi^2}$. This implies that if the latter condition on the dimension of $\chi$ is met, we can constrain the OPE coefficient $C_{\chi\O\chi}$ (by plugging in $E=E_\text{min}$ in \eqref{37})
\begin{align}
\ausricht
C_{\chi\O\chi}^2 &\lesssim 8\pi e \left(\frac{c}{12}\right)^{\frac{3}{4}}  \left(1-\frac{12\D_\chi}{c}\right)^{\D_\O-\frac{1}{2}}\frac{ \Gamma\left( \Delta_\O \right)^2}{\Gamma(2\Delta_\O)} \left(1-\sqrt{1-\frac{12\Delta_\chi}{c}}\right)\\
& \lesssim 4\pi e \left(\frac{12}{c}\right)^{\frac{1}{4}} \frac{ \Gamma\left( \Delta_\O \right)^2}{\Gamma(2\Delta_\O)}  \Delta_\chi+ \mathcal{O}(\D_\chi^2)\,.
\endeausricht
\end{align}
This provides a novel universal constraint on light data of a CFT. As it can be seen, the derivation uses modular bootstrap and the statistics that follow from it. This is  a surprising by-product of our analysis.

\section{Off-diagonal two-point functions}
\label{sec:off-diag-2-point-functions}
The analysis of the previous section provides results of the averaged OPE coefficients which are consistent with the ETH ansatz. In this section we shall work with a different probe and investigate the off-diagonal elements thereof. In what follows, the results fall under the context of the stronger form of ETH which posits that single energy eigenstates are thermal \cite{Garrison:2015lva}. The following calculation will however be confined to the large central charge regime.

The perturbative monodromy method can be used to derive an off-diagonal generalization of the heavy-light conformal block $\calv_P(h_{H_1},h_{H_2},h_L,c;z)$ in the large $c$ limit with all the scaled operator dimensions $h_i/c$ held fixed and set hierarchically. The heavy conformal dimensions ($h_{H_1},\, h_{H_2}$) much larger than the light ones $(h_L)$ and the exchanged primary $(h_P)$. 
We note that, in a different regime of parameter space, with $h_{H_1}/c$, $h_{H_2}/c$, $h_{H_1} - h_{H_2}$, $h_P$, $h_L$ fixed and $c \rightarrow \infty$, $\calv_P$ has been obtained from the global conformal block using a background field method in  \cite{Fitzpatrick:2015zha}. 

\subsection{Off-diagonal conformal blocks using the  monodromy method}\label{sec-mono}

We consider the correlation function two heavy and two light operators. 
The conformal partial wave expansion of this correlator is,
\beq
\langle H_2(\infty)  L(x,\bar{x}) L(y,\bar{y}) H_1(0) \rangle = \sum_P C_{H_1H_2P} C_{LLP} \calv_P(x,y,h_i) \bar{\calv}_P(\bar{x},\bar{y},\bar{h}_i).
\eeq
Here, $H_1$ and $H_2$ are heavy operators and, in general, $H_1\neq H_2$. Here we choose $H_1 > H_2$ without loss of generality. Note that by conformal transformations we could have put the insertions at $(x_1,x_2,x_3,x_4) = (0,x,1,\infty)$. However, since we shall like to make contact with a holographic interpretation it turns out to be transparent to keep the light operators at $x$ and $y$.
We shall fix our attention to the holomorphic part of the conformal block, $\calv_P(h_i,c;x,y)$. 
In the asymptotic heavy limit the states correspond to black hole geometry in the bulk where the back-reaction due to light operators is negligible. As mentioned earlier, the monodromy technique that we use \cite{Fitzpatrick:2014vua}\footnote{For readers interested in a more detailed description of the method than we present in the present text, we recommend the nice review in \cite{Fitzpatrick:2014vua} on it whose general logic we follow.} is valid only in the large central charge limit, whilst keeping the ratios $\epsilon_i \equiv {6}h_i/c$ fixed. In this limit, the blocks exponentiate
\beq 
\calv_P(x,y,h_i) \sim e^{-\frac{c}{6} f_P(\eps_i;x,y) }.\label{mono-block}
\eeq
The monodromy method considers a conformal block of interest with the presence of an auxiliary field $\hat{\psi}(z)$ which has a null state at level 2. The null state condition can be translated into the differential equation 
\beq \label{ode-1}
\partial_z^2\psi(z,z_i) + T(z,z_i) \psi(z) = 0, ~ ~~ T(z,z_i) = \sum_i \left\{ \frac{\epsilon_i}{(z-z_i)^2} - \frac{6}{c} \frac{c_i}{z-z_i}\right\}, 
\eeq
for $\psi(z,x,y)\equiv \vev{H_1(z_1)  L(z_2)L(z_3) \ket{p} \bra{p} \hat\psi(z)H_2(z_4)}$. Here $T(z,z_i)$ is the stress-tensor wavefunction in the presence of the heavy and light operator insertions.  The strategy is to solve for the function $f_P$ in \eqref{mono-block} by demanding consistent monodromies of the solutions of the above differential equation order by order in perturbation theory. The perturbative parameters in our case is in $\epsilon_L$, $\epsilon_P$ and $\epsilon_{H_1} - \epsilon_{H_2} \equiv 2 \epsilon$. Thus in addition to the exponentiation of block limits, we also work in the limit when, ${(\epsilon_{H_1} + \epsilon_{H_2})}/{2} \equiv \bar{\epsilon} \gg \epsilon_L, \epsilon, \epsilon_P$.  The $c_i$ are accessory parameters, $c_i = \partial_{z_i} f_P$. We also have the behaviour $T(z)\sim z^{-4}$ for large $z$, which imposes the constraints 
\beq \label{constraints}
\sum_i c_i = 0, ~~~ \sum_i c_i z_i - \epsilon_i = 0, ~~~ \sum_i c_i z_i^2 - 2 \epsilon_i z_i = 0\,. 
\eeq
These constraints allow us to re-express the stress-tensor wavefunction as, 
\beq\label{Tz}
T(z,z_i) = \frac{\bar{\epsilon} + \epsilon}{z^2} + \epsilon_L\left( \frac{1}{(z-x)^2} + \frac{1}{(z-y)^2}\right) + 2 \frac{\epsilon_L + \epsilon}{z(y-z)} - c_x \frac{x(x-y)}{z(z-x)(y-z) } .
\eeq
As mentioned earlier, in the heavy-light limit $\bar{\epsilon}$ is larger than other parameters, hence the zeroth order homogenous equation \eqref{ode-1}  is
\begin{equation}
\partial_z^2\psi^{(0)}  + \frac{\bar{\epsilon}}{z^2}\psi^{(0)} = 0\,,
\end{equation}
which has two solutions
$$
\psi^{(0)}_\pm  = z^{\frac{1 \pm \alpha_{\text{avg}} }{2} }\,,
$$
where $\alpha_{\text{avg}} = \sqrt{1 - 4 \bar{ \epsilon}} = \sqrt{ 1 - 12 ({h_{H_1} + h_{H_2}})/{c} }$.
The zeroth order solution can be used to find the perturbative solution at the first order, $\psi^{(1)}_\pm$, using the method of variation of parameters. This solution comes with a non-trivial monodromy. If we encricle $\psi^{(1)}_\pm(z)$ around $z_i = x$ and $y$, then the monodromy matrix, $M^{(1)}_\pm$ at the first order should satisfy
\beq
\tr M^{(1)}_\pm + \det M^{(1)}_\pm = 4 \pi^2 \epsilon_P^2\,.
\eeq
This equation can be solved for $c_x$ which is given by
\beq
x c_x  =  u^{\alpha_{\text{avg}}}\frac{(\epsilon + \epsilon_L - \alpha_{\text{avg}}\epsilon_L)}{u^{\alpha_{\text{avg}}} - 1} -\frac{(\epsilon + \epsilon_L +\alpha_{\text{avg}}\epsilon_L)}{u^{\alpha_{\text{avg}}} - 1} \pm \frac{\sqrt{ \epsilon^2 (u^{\alpha_{\text{avg}}} - 1)^2 + u^{\alpha_{\text{avg}}} \epsilon_P^2 \alpha_{\text{avg}}^2 }}{u^{\alpha_{\text{avg}}} - 1}  .\label{xcx}
\eeq
where we defined, $u= y/x$. 
Note that in this case we have two non-trivial accessory parameters, $c_x = \partial_x f_P$ and $c_y = \partial_y f_P$. The constraints \eqref{constraints} yield
\begin{align}
 y\partial_y f_P + x \partial_x f_P &= 2( \epsilon+\epsilon_L ).
\end{align}
The general solution of the above is
\beq
f_P = 2( \epsilon+\epsilon_L )\log x + g(u).
\eeq
Now differentiating the above with respect to $\log x$ we obtain
\beq
x \partial_x f_P = 2(\epsilon + \epsilon_L) - u g'(u).\label{xcx-int}
\eeq 
This object by definition is just $x c_x$ which is a function of $u$ \eqref{xcx}. Therefore we can now integrate \eqref{xcx-int} to find $g(u)$. The root in \eqref{xcx} and the constant of integration is fixed so as to yield the expected OPE limit of the conformal block. Putting together everything, we can bring the result to the following form (till the linear order $\epsilon,\epsilon_L,\epsilon_P$)
\begin{equation}
f_P^{H_1,H_2,L}(x,y) =  \epsilon \log x y + 2\epsilon_L \log \frac{x^{\alphaa} - y^{\alphaa}}{\alphaa ( x y)^{\frac{\alphaa-1}{2}}} + \epsilon_P \log \frac{\alphaa(x^{\alphaa/2}+y^{\alphaa/2})}{4(x^{\alphaa/2}-y^{\alphaa/2})}.\label{mono-block-f}
\end{equation}
In the diagonal limit, $\epsilon = 0$ and the operator locations $x \rightarrow 1$ and $y \rightarrow x$ the answer reduces to \cite[equation (2.26)]{Hijano:2015rla}. The conformal block can then be obtained from the above function by exponentiation \eqref{mono-block}
\begin{align}\label{exp-block}
\cV_P(x,y) = (xy)^{-\frac{h_{H_1}-h_{H_2}}{2}} \left[ \frac{\alphaa ( x y)^{\frac{\alphaa-1}{2}}}{x^{\alphaa} - y^{\alphaa}} \right]^{2h_L} \left[\frac{4(x^{\alphaa/2}-y^{\alphaa/2})}{\alphaa(x^{\alphaa/2}+y^{\alphaa/2})}\right]^{h_p}\, , 
\end{align}
which is the main result of this section. 
For a sparse spectrum of light operators, this result provides an analytic expression of the smooth function $f_{LL}(\bar{\epsilon}, \epsilon)$ which appears in  the version of the ETH ansatz \eqref{eth-ansatze} for the bilocal probe  $L(x)L(y)$. We notice that there is a power law decay $(xy)^{- ({h_{H_1}- h_{H_2}})/{2} }$ for the off-diagonals. The full 4-point function also appears with the OPE coefficients $C_{LL\chi}C_{{H_1}{H_2}\chi}$. Here, $\chi$ is the lightest primary appearing in the fusion of $H_{1}$ and $H_2$ and two $L$ operators. For a typical state, our estimate of the $C_{{H_1}{H_2}\chi}$ from Section \ref{sec:off-diag-1-point} shows the presence of entropic suppression also in this case. Note that, unlike the diagonal heavy-light correlator (in a typical holographic CFT), the dominant contribution is not from the vacuum block in the off-diagonal case. If the light spectrum is sufficiently sparse, the dominant contribution to $\vev{H_2 LL H_1}$ is from the block of the light primary $\chi$.

The parameter $\alphaa$ becomes imaginary when $(h_{H_1}+h_{H_2})/2 > c/24$. In the bulk dual this corresponds to the black hole regime. The conformal block \eqref{exp-block} then acquires periodicities akin to thermal correlator. In particular, the second factor $[\cdots]^{2h_L}$ in \eqref{exp-block} has the form of $[\sinh(\pi \ell/\beta)]^{-2h_L}$, when written in cylinder coordinates. Here, $\beta$ is exactly the same as the saddle which we found in the modular bootstrap analysis -- equation \eqref{beta-saddle}. The frequency $\omega$ is $12\epsilon/c =h_{H_1} - h_{H_2}$. 

\subsection{Comparison with other approaches}\label{sec-comp}

\subsubsection*{Zamolodchikov recursion}
The Virasoro blocks can be computed at any value of central charge but are not known in closed form \cite{Zamolodchikov:1985wn,Perlmutter:2015iya}. Nevertheless, the coefficients of the block in a small cross-ratio expansion can be determined algorithmically using Zamolodichikov's recursion\footnote{ This is formally a series in $q$ which is related to the cross-ratio $z$ via, $q = e^{i \pi \tau},  ~~ \tau = i \frac{K(1-z)}{K(z)}$.}. 

The S-channel (or $z\to 1$) expansion of $\log \cV_{h_p}(c,h_L,h_{H_1},h_{H_2},z)$ can be organized in the following manner
\begin{align}
\log \cV_{P}(c,h_L,h_{H_1},h_{H_2},z)= (h_p -2h_L)\log (1-z) + \sum_{n=0}^\infty v_n (1-z)^n 
\end{align}
In the regime of parameter space of the monodromy analysis, we have the following expansions for the coefficients $v_n$
\begin{align}
\ausricht
v_0 &= \frac{h_{H_2}-h_{H_1}+h_p}{2}   , \\
v_1 &= \frac{h_{H_2}-h_{H_1}}{4} +\frac{h_{H_1}+h_{H_2}}{c}h_L + \left(\frac{3}{16}+\frac{h_{H_1}+h_{H_2}}{4c} \right) {h_p} + \cdots ,\\
v_2 &= \frac{h_{H_2}-h_{H_1}}{6} +\frac{h_{H_1}+h_{H_2}}{c}h_L + \left(\frac{5}{48}+\frac{h_{H_1}+h_{H_2}}{4c} \right) {h_p} + \cdots ,\\
v_3 &= \frac{h_{H_2}-h_{H_1}}{8} +\frac{9(h_{H_1}+h_{H_2})}{10c}h_L + \left(\frac{35}{512}+\frac{71(h_{H_1}+h_{H_2})}{320c} \right) {h_p} + \cdots .
\endeausricht 
\end{align}
Note that the above expansion is obtained by performing a $1/c$ expansion first. This is followed by a scaling $h_p/c \to \delta\, h_p/c $, $h_L/c \to \delta\, h_L/c $ and $(h_{H_1}-h_{H_2})/c \to \delta\, (h_{H_1}-h_{H_2})/c$ and then an expansion to the linear order in $\delta$. 
These coefficients match exactly with those of the monodromy method \eqref{exp-block}.

It is worthwhile to observe that there are divergences arising from thermal periodicities in the conformal block calculated using the monodromy method. This happens for $h_{H_1}+h_{H_2}>c/12$ for which $\alpha_\text{avg}$ is imaginary. These divergences are often referred to as forbidden singularities. As shown in \cite{Chen:2017yze}, this is purely an artefact of the large central charge limit\footnote{As an intermediate step there is also a finite $c$ resolution obtained by resumming the ${\mathcal O}(h_L/c)$ effects in the monodromy method \cite{Faulkner:2017hll}. This gives rise to ``forbidden-branch-cuts" that resolve the forbidden singularities, while still not altering ETH expectations.}. The blocks computed numerically using the Zamolodchikov recursion relations do not display these divergences. We have checked that the story is the same for the off-diagonal blocks.

\subsubsection*{Heavy-light blocks from the background field method}
At large central charge, the heavy light conformal block has also been investigated in the regime in which the dimensions of the light and  intermediate operators and differences between dimensions of heavy operators held fixed \cite{Fitzpatrick:2015zha}. This is a  different regime in the parameter space of conformal dimensions from the one covered by the monodromy method. The monodromy method has the ratios of the conformal dimensions and the central charge held fixed. The result of \cite{Fitzpatrick:2015zha} is 
\begin{align} \label{glo-block}
\cV_P (z) \simeq (z)^{h_L (\alphaa -1 )} \left( \frac{1-z^{\alphaa} }{\alphaa}\right)^{h_p -2 h_L } {}_2F_1\left( h_p - \tfrac{(h_{H_2}-h_{H_1})}{2\alphaa}, h_p, 2h_p, 1-z^{\alphaa} \right) 
\end{align}
which also probes the off-diagonal sector. It can be shown to agree with the expansion using the Zamolodchikov recursion but using a different sequence of limits. One needs to expand first in $h_L$, $h_p$ and $h_{H_1}-h_{H_2}$ and then expand in $1/c$. This result also remarkably agrees on the nose with a bulk computation \cite{Hijano:2015qja}. 

It is interesting to contrast the block \eqref{glo-block} with the monodromy block \eqref{exp-block}. 
In order to do this, we use the following Euler identity\footnote{It reads ${}_2 F_1 (a,b,c;z) = (1-z)^{c-a-b}{}_2 F_1 (c-a,c-b,c;z)$ in the original form.}  for the hypergeometric factor
\begin{align}
{}_2F_1\left( h_p - \tfrac{(h_{H_2}-h_{H_1})}{2\alphaa}, h_p, 2h_p, 1\!-\!z^{\alphaa} \right) = z^\frac{h_{H_1} -h_{H_2} }{ 2 } {}_2F_1\left( h_p + \tfrac{(h_{H_2}-h_{H_1})}{2\alphaa}, h_p, 2h_p, 1\!-\!z^{\alphaa} \right). \nn 
\end{align}
Plugging this in \eqref{glo-block} this gives
\begin{align}
\cV_P (z) \simeq&  \ z^\frac{h_{H_2} -h_{H_1} }{ 2 } \left( \frac{\alphaa \, z^\frac{\alphaa -1 }{ 2} }{1-z^{\alphaa} }\right)^{2 h_L }   \nn \\ 
&\ \times\left( \frac{1-z^{\alphaa} }{\alphaa}\right)^{ h_p } {}_2F_1\left( h_p + \tfrac{(h_{H_2}-h_{H_1})}{2\alphaa}, h_p, 2h_p, 1-z^{\alphaa} \right).
\end{align}
The factors on the first line precisely agree with the $h_p$-independent piece of the monodromy block \eqref{exp-block}. The piece that is dependent on the intermediate conformal dimension differs. Both the results agree for diagonal and vacuum case, $h_{H_1}=h_{H_2}$ and $h_p=0$. 
\begin{figure}[b!]	
  \centering
  \includegraphics[width=7cm]{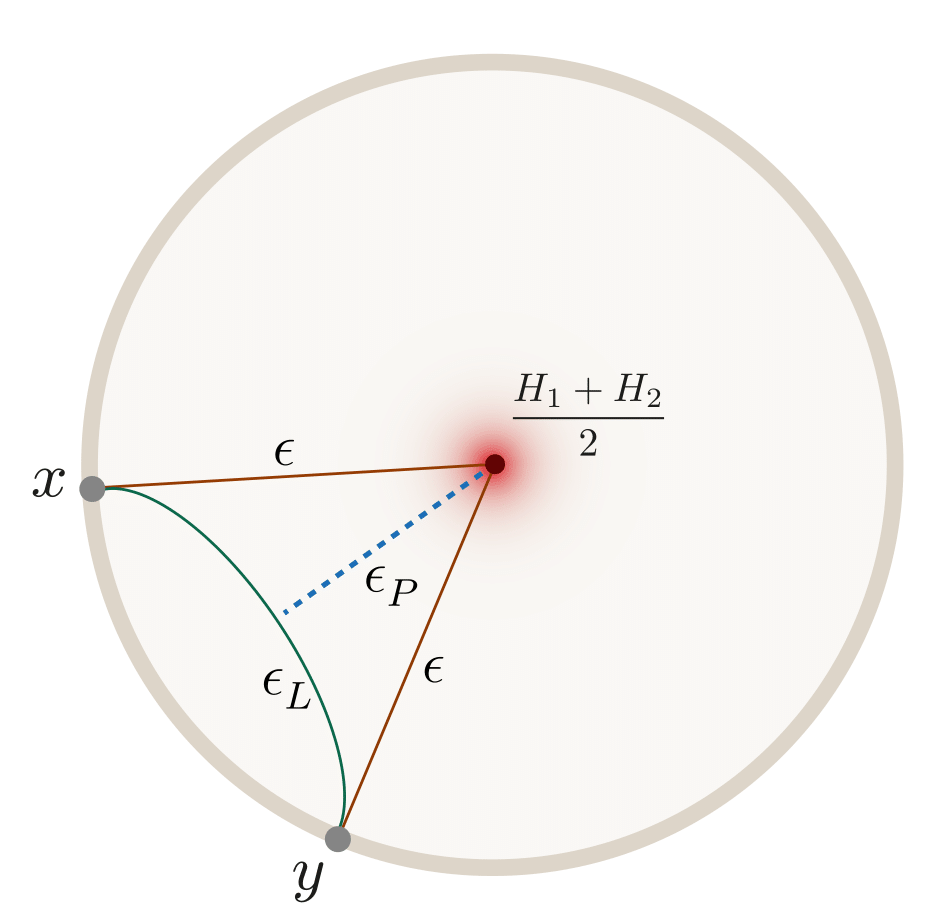}
  \caption{Geodesic configuration at the $AdS_3$ time slice equivalent to the semi-classical Virasoro blocks result \eqref{exp-block} calculated using the monodromy method.}\label{geodesics}
\end{figure}

\subsubsection*{Holography}
 The result for the conformal block using the monodromy method \eqref{mono-block-f} can be interpreted in the bulk  dual via a configuration of geodesics on a fixed time slice of the global $AdS_3$ in presence of a conical singularity at the origin \cite{Hijano:2015rla}. The metric is given by
 \begin{align}
 ds^2 = \frac{\alphaa^2}{\cos^2\rho} \left[-dt^2 +\frac{1}{\alphaa^2}d\rho^2 +\sin^2 \rho d\phi^2 \right], \ \ \text{with } \alphaa = \sqrt{1-12(h_{H_1}+h_{H_2})/c}.
 \end{align}
 This is depicted in Fig.~\ref{geodesics}. The conical-deficit background can be thought of as created by a heavy scalar with holomorphic conformal dimension given by the average of those of $h_{H_1}$ and $h_{H_2}$. The dual CFT then lives on a cylinder. 
 
The operators dual to the light scalar $L$ and the one that labels the block $P$ are shown in Fig.~\ref{geodesics} by the green curve and the blue dotted line respectively. Both of them correspond to massive scalar fields, $\phi_L$ and $\phi_P$ in the bulk with mass given by $m_L= 2\sqrt{h_L(h_L-1)}$ and similarly for $m_P$. The last term in equation  \eqref{mono-block-f} is precisely  the geodesic length of $\phi_P$, while the second term is the geodesic length of $\phi_L$ times conformal factors coming from the transformation from the cylinder to the plane. The remaining term (\ie the first one in \eqref{mono-block-f}) is the product of regularized radial geodesic lengths, from the boundary anchoring points of $\phi_L$ to the centre.\footnote{The interested reader can find the computation of these geodesic lengths in   \cite{Banerjee:2016qca}.} In the block both these come weighted with the same power of $h_{H_1} -h_{{H}_2}$. This is interpreted as yet another bulk massive scalar field, $\phi_\epsilon$ which is dual to an operator carrying conformal dimension given by the difference of the dimensions of the heavy operators.\footnote{If we had instead done perturbation in $h_{H_1} - a h_{H_2}$ with $a \in (0,1)$, this would be reflected in the monodromy solution and in the corresponding bulk interpretation as a shift in the background and in the masses of $\phi_\epsilon$.} 

\section{Conclusions}\label{sec:conclusions}

In this work we have estimated the asymptotic behaviour of averaged matrix elements of an arbitrary scalar operator in the energy eigenbasis in the context of 2$d$ CFTs. The off-diagonal matrix elements are universal and  consistent with the eigenstate thermalization hypothesis. They dictate quantum fluctuations and, hence, play a key role in the investigation of non-equilibrium physics. Our analysis, which is valid for any positive value of the central charge, involves a coarse graining over exponentially large number of states. The final result for $\overline{|\langle E | \O | E' \rangle|^2}$ for large $E,E'$ with $E-E'$ small and non-zero is of the form $e^{-S} \Gamma$, where $S$ is the entropy at the average energy and $\Gamma$, which is written in terms of gamma functions, measures the spread of the fluctuations. We have also found the asymptotics of actual OPE coefficients corresponding to primaries using properties of 2-point torus blocks. These findings from modular bootstrap also provide information that allows to investigate late time behaviour of thermal 2-point functions on the torus.  This can potentially furnish a 2$d$ CFT analogue of a part of the analysis performed for the SYK model in \cite[\S 7]{Cotler:2016fpe}. 

We then used our results along with results from \cite{km} to provide a constraint using positivity of the variance of diagonal matrix elements. While we were able to give an upper bound to the variance of the diagonal matrix elements, it may be possible to find the diagonal variance itself. This will require estimation of a weighted sum of the form $\sum'_{ij}  |\vev{ E_i | A | E_j }|^2 \delta(E_i - E) \delta(E_j - E) $, where the prime indicates that we sum over non-identical states which are degenerate in their energies. 

In the latter half of this work we studied off-diagonal Virasoro  blocks of the form, $\vev{H_1 L L H_2}$ in the large central charge limit. The calculation was done using the monodromy method where the conformal dimensions were taken to be of the order of the central charge. We performed the calculation with the conformal dimensions of both $H_1$ and $H_2$ much larger than those of $P$ (intermediate exchange operator), $L$, and the difference in the dimensions of $H_1$ and $H_2$. We obtained the answer for the block till linear order in $h_L/c, (h_{H_1}\!-h_{H_2})/c$ and $h_P/c$. The answer for the blocks bears some thermal features. Correspondingly, the holographic interpretation involves geodesics in a conical defect geometry. The scalar probes forming the geodesic network have masses proportional to $h_L$, $h_P$ and $h_{H_1}\!-h_{H_2}$. The conical defect geometry can be analytically continued to a BTZ black hole geometry whose temperature is once again dictated by the average, $(h_{H_1}\!+ h_{H_2})/2$. In this case the thermal description emerges, not by coarse-graining, but at large central charge, \ie in semi-classical limit.  

There are various possible generalizations and applications of this work. It is  interesting to look at the off-diagonal out of time ordered correlator of the form, $\la W(t) V W'(t) V \ra_\beta$ in a thermal state, when the conformal dimensions of $W$ and $W'$ are only slightly different. Upon following the relevant analytic continuation prescription of the Euclidean correlator \cite{Roberts:2014ifa} one obtains for the OTOC, 
\beq
\la W(t+i\epsilon_1) V(i \epsilon_3) W'(t+ i \epsilon_2) V(i\epsilon_4) \ra_\beta \approx \left( \frac{1}{ 1 - \frac{ 12 \pi i (h_W + h_{W'}) }{\epsilon_{12}^* \epsilon_{34}} } e^{(2\pi/\beta)(t-t_*-x)}\right)^{2 h_v } . \nn 
\eeq
The dual interpretation is that of correlators of the light scalars dual to $V$ in a shockwave background corresponding to the averaged weight of $W$ and $W'$. It would be interesting to investigate further implications of this. Furthermore, it would be worthwhile to explore a generalization of black hole collapse with a non-uniform distribution of matter by a continuum version of our monodromy calculation \cite{Anous:2016kss}.

If ETH holds approximately in a local theory for all operators within a subregion $A$ in the excited energy eigenstate, $\ket{\psi}$, then the reduced density matrix satisfies,
$$
\rho_A^\psi = \Tr_{A^c} \ket{\psi} \bra{\psi} \approx \Tr_{A^c} e^{-\beta H} \simeq e^{-\beta H_A},
$$
where $\beta$ is determined from $\vev{H}_\beta = \langle \psi| H | \psi \rangle$ and $H_A$ is the Hamiltonian restricted to the subregion $A$. The above equality can be used to approximately construct the full Hamiltonian given an eigenstate that satisfies ETH \cite{Garrison:2015lva, 2017arXiv171201850Q}.  { Once  the conditions for validity of ETH for CFTs are well  understood, the construction of a chaotic local Hamiltonian can be potentially possible, with a view towards building holographic CFTs from bottom-up. }

If the conformal field theory has additional conserved currents, then the ensemble approximating of charged excited states is the grand canonical ensemble. The modular properties get modified when there is an additional Kac-Moody along with the Virasoro as the chiral algebra \cite{Benjamin:2016fhe}. This gets reflected in the modular bootstrap analysis of diagonal OPE coefficients \cite{Das:2017vej} by the appearance of {spectral-flow invariants} and Ahranov-Bohm like phases. It will be interesting to see how the presence of additional global symmetries modify the off-diagonal matrix elements and, in particular, how the fluctuations manifestly depend on the conserved charges.

\section*{Acknowledgements}
It is a pleasure to thank Jan de Boer, Sumit Das, Matthias Gaberdiel, Tarun Grover,  Tom Hartman, John McGreevy, Sridip Pal and especially Per Kraus for fruitful discussions. 
DD acknowledges the support provided by the Alexander von Humboldt Foundation and the Federal Ministry for Education and Research through the Sofja Kovalevskaja Award.  
The work of SD is supported by the NCCR SwissMAP, funded by the Swiss National Science Foundation. SD also thanks AEI Potsdam and Swansea University for hospitality where parts of this work were completed. Finally, the authors thank the participants and organizers of `Strings, Geometry \& Black Holes' at King's College London for simulating discussions and an opportunity to present this work.

\appendix

\section{Mellin-Barnes integrals}\label{mellin}
The following integral is useful for some of the calculations of this work. For any complex $A$ and $B$
\begin{align}
\int_{-i\infty}^{i\infty}  \frac{ds}{2\pi i} \left(\frac B A\right)^s \Gamma(p-s) \Gamma(q+s) = \frac{ \Gamma(p+q) B^p A^q }{ (A+B)^{p+q}}.
\end{align}
This is of the form a Fourier transform of the product of  Gamma functions if $A=e^{-\pi i\om}= -B^{-1}$.
A special case of the inverse of the above relation is 
\begin{align}\label{inv-MB}
\hspace*{-0.2cm}\int_{-\infty}^{\infty} dt \, e^{-i\om t} \left[\frac{\pi /\beta }{ \sinh(\tfrac{\pi t}{\beta}\pm i \epsilon)}\right]^{2\D_\O} =  (-1)^{\mp \Delta_\O} \left(\frac{2\pi}{\beta}\right)^{2\D_\O-1} e^{\mp \frac{\beta \om }2 } \frac{\left|\Gamma \left(\D_\O+ i\frac{\beta \om }{ 2 \pi} \right)  \right|^2}{\Gamma(2\D_\O)} \,
\end{align}
The $\pm i\epsilon$ prescription is chosen in order to avoid the singularities along the real $t$ axis. 

\begin{footnotesize}
\bibliography{refs}
\bibliographystyle{bibstyle2017}
\end{footnotesize}
\vspace{2cm}

\end{document}